\documentclass{article}
\usepackage[T1]{fontenc} % add special characters (e.g., umlaute)
\usepackage[utf8]{inputenc} % set utf-8 as default input encoding
\usepackage{ismir,amsmath,cite,url}
\usepackage{graphicx}
\usepackage{color}
\usepackage{todonotes}

\usepackage{graphicx}  %Required
\usepackage{booktabs} % For formal tables
\usepackage{caption}
\usepackage{subcaption}
\usepackage{enumerate}
\usepackage{multirow}

\newcommand{\algname}[1] {{\fontfamily{cmtt}\selectfont {#1}}}

\title{The Relationship Between the Consistency of Users' Ratings and Recommendation Calibration}

\multauthor
{Masoud Mansoury$^1$ \hspace{0.75cm} Himan Abdollahpouri$^2$ \hspace{0.75cm} Joris Rombouts$^1$ \hspace{0.75cm} Mykola Pechenizkiy$^1$} { \bfseries{}\\
  $^1$ Eindhoven University of Technology, the Netherlands\\
  $^2$ University of Colorado Boulder, USA\\
  {\tt\footnotesize m.mansoury@tue.nl, himan.abdollahpouri@colorado.edu, j.c.rombouts@student.tue.nl, m.pechenizkiy@tue.nl}
}
\def\authorname{Masoud Mansoury, Himan Abdollahpouri, Joris Rombouts, Mykola Pechenizkiy}

\begin{document}

\maketitle
\begin{abstract}
%\todo[inline]{Do Users with More Informative Profile Receive Fairer Recommendation?}
%\todo[inline]{Does Having a More Informative Profile Ensures Receiving More Calibrated Recommendations?}
Fairness in recommender systems has recently received attention from researchers. Unfair recommendations have negative impact on the effectiveness of recommender systems as it may degrade users' satisfaction, loyalty, and at worst, it can lead to or perpetuate undesirable social dynamics. One of the factors that may impact fairness is \textit{calibration}, the degree to which users' preferences on various item categories are reflected in the recommendations they receive. 

The ability of a recommendation algorithm for generating effective recommendations may depend on the meaningfulness of the input data and the amount of information available in users' profile. In this paper, we aim to explore the relationship between the consistency of users' ratings behavior and the degree of calibrated recommendations they receive. We conduct our analysis on different groups of users based on the consistency of their ratings. Our experimental results on a movie dataset and several recommendation algorithms show that there is a positive correlation between the consistency of users' ratings behavior and the degree of calibration in their recommendations, meaning that user groups with higher inconsistency in their ratings receive less calibrated recommendations.

%Depending on the amount of information available in users' profile, recommendation algorithms may generate specific degree of calibrated recommendations for different groups of users.   

%One of the key factors on the performance of recommender systems is the amount of information that is available in a target user's profile for a recommendation algorithm when generating recommendations for that user. In this paper, we aim to explore the relationship between the informativeness of users' profile and fairness of recommendations that they might receive. We conduct our analysis on different groups of users based on the informativeness of their profile. Our experimental results on a movie dataset and four well-known recommendation algorithms that there is positive correlation between informativeness of profiles and fairness of recommendations.

 %In some cases biases in the original data maybe amplified or reversed by the underlying recommendation algo-rithm. In this paper, we explore how different recommendationalgorithms reflect the tradeoff between ranking quality and bias dis-parity. Our experiments include neighborhood-based, model-based,and trust-aware recommendation algorithms

%In this paper, we analyze users' profile and recommendation results to understand the possible reasons behind unfair recommendations. We investigate the relationship between the richness of users' profile and the degree of fairness in recommendation results they might receive. We study this relationship on different segments of users based on their profile richness.
\end{abstract}
\section{Introduction}

Recommender systems are powerful tools for predicting users' preferences and generating personalized recommendations. These systems, while effective, often suffer from lack of fairness in recommendation results, meaning that the outputs of recommendation algorithms are, in some cases, biased against some protected groups \cite{ekstrand2018}. As a result, this discrimination among users will negatively affect users' satisfaction, loyalty, and overall effectiveness of the system.

Unfair recommendation is often defined as the situation that a recommendation algorithm behaves differently when generating recommendations for different groups of users (i.e., protected and unprotected groups). As an example, when users who belong to the unprotected group receive more accurate recommendations than the users in the protected group, we say there is discrimination against the protected group. This unfair behavior can originate from either the underlying biases in the input data used for training \cite{burke2017,virginia2018,abdollahpouriWSDM} or the result of recommendation algorithms \cite{yao2017}. 

Abdollahpouri et al. in \cite{abdollahpourirmse1} showed that popularity bias has a negative impact on the fairness of recommendation outputs. In that work, authors showed that the recommendations generated for the majority of users are concentrated on popular items even for those who are interested in long-tail and non-popular items. A more similar analysis to our work is done in \cite{abdollahpouriWSDM} where authors showed how popularity bias is correlated with the miscalibration of the recommendations and how different user groups with varying degree of interest in popular items experience different levels of miscalibration. 

In this paper, we aim to do more exploration on the possible reasons for discrimination in recommendation results. Our hypothesis is that the richness of a user's profile might have impact on how the algorithm performs for that user. To explore this, we analyze users' profile and investigate the relationship between the consistency of users' ratings and the degree of calibrated recommendations. We believe that the lack of consistency in user's profile can be one possible reason for miscalibrated recommendations as recommender system is unable to correctly predict user's preferences. We discuss the approach for measuring profile consistency in next section. 
%This is important to be studied because it helps to understand the reason behind unfair recommendations. 

%Our experimental results show that profile richness is positively correlated with calibrated recommendation meaning that richer profiles (i.e., more informative profiles) will receive more calibrated recommendations.

\section{Profile consistency}

We define a rating to be consistent if it is in agreement with the ratings given by other users. For instance, if a user has given 5 to an item with the average rating of 2, it means his rating has an inconsistency of degree 3. Profile consistency refers to the fact that how similar a user rates an item compared to the majority of other users who have rated that item. This has been referred to the gray sheep problem in the literature \cite{ghazanfar2014leveraging}. Since collaborative filtering approaches use opinions of other users (e.g., similar users) for generating recommendations for a target user, it is highly possible that inconsistent profiles do not receive effective recommendations. Given a target user, $u$, and $I_{u}$ as all items rated by $u$, inconsistency of $u$ can be calculated as:

\begin{equation} \label{eq:consistency}
\begin{aligned}
inconsistency_{u}=\frac{\sum_{i \in I_{u}}|r_{u,i}-\overline{r_{i}}|}{N_{u}}
\end{aligned}
\end{equation}

where $r_{u,i}$ is the rating provided by $u$ on item $i$, $\overline{r_{i}}$ is average of ratings assigned to item $i$, and $N_{u}$ is the number of items rated by $u$.

\section{Calibration measure}

Measuring fairness of recommendation results is a complex task. Several metrics have been recently proposed for measuring the equity of recommendation results \cite{burke2017,yao2017,virginia2018}. Bias disparity \cite{virginia2018,mansoury2019biasdisparity} is one of those metrics that measures how much an individual's recommendation list deviates from his or her original preferences in the training set across an item's category. The issue with bias disparity is that it calculates the bias value for a group of users on a specific item category and does not return the overall bias value for a group of users across all item categories.

Calibration of recommendations is another factor that affects fairness in recommender systems \cite{steck2018}. Calibration measures the distance between users' preferences in training data and the predicted preferences in recommendation lists. Distance equals to zero indicates perfect calibration, while distance larger than zero indicates a degree of \textit{miscalibration}. For the rest of the paper, we use the term miscalibration to refer to this distance value. 

Original preferences in train set and predicted preferences in recommendation lists are represented as distributions across item categories and the distance between these two distributions shows the degree of miscalibration. The main incentive behind having calibrated recommendation is the fact that recommendation lists should appropriately represent users' profile/interest in train data. Assume a user's profile consists of 70\% action movies and 30\% adventure movies. Then, it is expected that the recommendation list for this user also contains the same proportion of each genre. 

For calculating the miscalibration, we follow the equations introduced in \cite{steck2018}. Given the distribution of items' category in user $u$'s profile as $p$ and the distribution of items' category in recommendation list generated for user $u$ as $q$, we use Kullback-Leibler divergence measure for calculating the distance between these two distributions for user $u$ as follow:

\begin{equation} \label{eq:kld}
\begin{aligned}
KL_{u}(p|\widetilde{q})=\sum_{c \in C}{p_{c}  log\frac{p_{c}}{\widetilde{q}_{c}}}
\end{aligned}
\end{equation}

where $C$ is item categories (e.g., genres in movie recommendations) and $\widetilde{q}$ is approximately similar to $q$ calculated as:

\begin{equation} \label{eq:q}
\begin{aligned}
\widetilde{q}_{c}=(1-\alpha).q_{c} + \alpha.p_{c}
\end{aligned}
\end{equation}

The purpose of $\widetilde{q}$ is to overcome the issue of zero values for some categories in $q$. Small value for $\alpha>0$ guarantees $\widetilde{q} \approx q$. In our experiments, we use $\alpha=0.01$ as suggested in \cite{steck2018}. 
%Since KL-divergence measures the distance between $p$ and $q$, lower KL value shows better results.

\captionsetup[table]{skip=4pt}
\begin{table}[t!]
\small
\centering
\captionof{table}{Accuracy of recommendation algorithms} \label{tab:accuracy}
\begin{tabular}{l|llll}
\toprule
 algorithm & \algname{UserkNN} & \algname{ItemkNN} & \algname{SVD++} & \algname{ListRankMF} \\
 \midrule
 precision@10 & 0.214 & 0.223 & 0.122 & 0.148 \\
 \bottomrule
\end{tabular}
\end{table}

\section{Experiments}

For experiments, we use MovieLens 1M (ML1M) dataset which is a movie rating data collected by the MovieLens\footnote{https://grouplens.org/datasets/movielens/} research group. In this dataset, 6,040 users provided 1,000,209 ratings on 3,706 movies. The ratings are in the range of 1-5 and the density of the dataset is 4.468\%. Also, each movie is assigned several genres. Overall, there are 18 genres in this dataset.

For performing experiments, we divided the dataset into train and test sets as 80\% and 20\%, respectively. The train set is used for building the model, and in the test condition, we generate recommendation lists of size 10 for each user.

After recommendation generation, for each user, we calculate a value for inconsistency of profile and a value for miscalibration. We measure inconsistency of profile using equation \ref{eq:consistency} and miscalibration of recommendations generated for a user using equation \ref{eq:kld}. For the purpose of presentation, we sort users based on their profile inconsistency and then group them into several groups with the same range. Finally, for each group we calculate the average of profile inconsistency and miscalibration.

Our experiments includes user-based collaborative filtering (\algname{UserKNN}), item-based collaborative filtering (\algname{ItemKNN}), singular value decomposition (\algname{SVD++}), and list-wise matrix factorization (\algname{ListRankMF}).
All recommendation models are optimised using Grid Search over hyperparameters and best results in terms of precision are reported here. Table \ref{tab:accuracy} shows the accuracy of those recommendation algorithms. We used \textit{librec-auto} and LibRec 2.0 for all experiments \cite{mansoury2018automating,Guo2015}.

\begin{figure*}[htp]
  \centering
  \begin{subfigure}[b]{0.24\textwidth}
        \includegraphics[width=\textwidth]{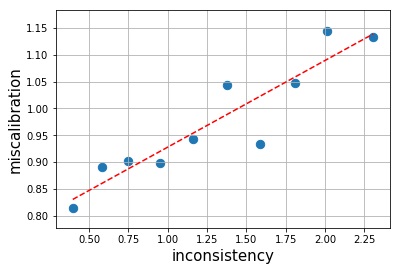}
        \caption{\algname{UserKNN}} \label{fig:corr:userknn}
  \end{subfigure}
  \begin{subfigure}[b]{0.24\textwidth}
        \includegraphics[width=\textwidth]{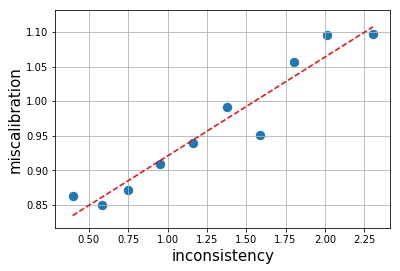}
        \caption{\algname{ItemKNN}} \label{fig:corr:itemknn}
  \end{subfigure}%
  \begin{subfigure}[b]{0.24\textwidth}
        \includegraphics[width=\textwidth]{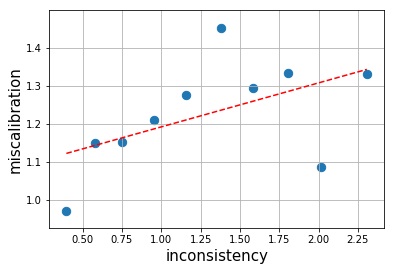}
        \caption{\algname{SVD++}} \label{fig:corr:svdpp}
  \end{subfigure}%
  \begin{subfigure}[b]{0.24\textwidth}
        \includegraphics[width=\textwidth]{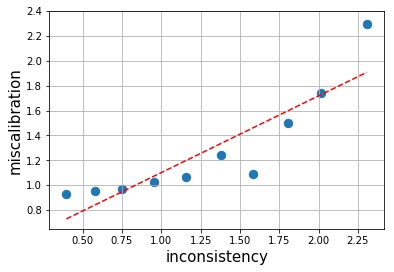}
        \caption{\algname{ListRankMF}} \label{fig:corr:svdpp}
  \end{subfigure}%
\caption{Relationship between inconsistency of users' profile and miscalibration of recommendations generated for those users.} \label{fig:corr}
\end{figure*}

\subsection{Experimental results}

%In this section, we present the results of our experiments on ML1M. Our experiments explore the relationship between users' profile richness and recommendation fairness (i.e., calibration). 

Figure \ref{fig:corr} shows the relationship between inconsistency in users' profiles and the miscalibration of the recommendations for each group. For all recommendation algorithms, there is a %\todo[inline]{Since you use the word strong you need to report R-squared} 
positive correlation between inconsistency of the ratings in the profile and miscalibration: as inconsistency increases, miscalibration will also increase. Except for \algname{SVD++}, there is a strong correlation for all other recommendation algorithms.

The correlation coefficient for \algname{UserKNN} is 93\%, for \algname{ItemKNN} is 96\%, for \algname{SVD++} is 53\%, and for \algname{ListRankMF} is 88\% which are indicative of strong correlation between inconsistency of profile and miscalibration, except for \algname{SVD++}. %\todo[inline]{The SVD++ does not show a strong correlation which is actually expected. UserKNN and ItemKNN are both neighborhood-based algorithms and therefore the consistency of user's profile ratings with the rest of the users is more important in these two algorithms}. 

These are interesting results as they show that users who provide inconsistent ratings will less likely receive calibrated recommendations. This can increase unfair situation such that different users will receive different level of calibration in their recommendation lists. Therefore, taking into account the inconsistency of users' profile when generating recommendations can alleviate unfairness of recommendation outputs.

\section{Discussion}

%Although in this paper we considered consistency of users' profile as the informativeness of those profiles for recommender systems, there are other ways of computing the informativeness of users' profile. 
Although in this paper we considered consistency of users' profile as a factor that has positive impact on the effectiveness of recommender systems, there might be other factors that also contribute to the effectiveness of these systems.

Profile size can be one of the factors for generating successful recommendations and may affect the performance of recommender systems. Users with low profile size or insufficient number of ratings are often known as \textit{cold-start users}. It has been long noted that these profiles are the source of concern for recommender systems as recommendation algorithms are unable to accurately predict their preferences \cite{lika2014}.

%We examine profile richness in terms of number of ratings provided by a user and entropy of the profile. A profile with large number of ratings will have positive impact on recommendation performance as it provides enough information about user preferences. On the other hand, entropy is a measure of amount of information provided by a profile. Thus, high entropy will positively affect recommendation performance.

Information gain (i.e. entropy) is one form of measuring informativeness of a profile and another factor that may affect the performance of recommender systems. A Profile with high entropy is the one where the user has provided ratings to a wide range of items from least preferred to most preferred ones. These profiles are informative because they provide both positive and negative feedback and recommender system will better learn to what recommend and what not recommend. We will consider aforementioned metrics (or combination of those metrics) for measuring informativeness or richness of a profile as our future work. 

%Note that the combination of profile size and entropy may better represent the profile richness. It is possible that a profile with large number of ratings do not provide much information about user preferences. Imagine a large profile with plenty of identical ratings. Although this profile has large number of ratings, those ratings do not very well present a user preferences as all ratings are similar. On the other hand, a small profile may show high entropy for the situation that different rating values (low and high ratings) are assigned to few items. Therefore, taking into account both profile size and entropy may avoid recognizing a less informative profile as informative one.

Our experiments in this paper are performed on a user-item rating data in movie domain. However, it can be extended to other datasets from different domains. In particular, as a future work, we intend to extend this work to music recommendation. We are interested in investigating whether inconsistency of a user's profile has any connection with the fact that some users have a niche taste and they might rate some popular songs differently from the majority of other users.

\section{Conclusion}

In this paper, we explored the relationship between the consistency of users' profile and calibration of recommendations. 
%Consistency of profiles in ratings and calibration of recommendations are used as measures of profile richness and recommendation fairness, respectively. 
Our experimental results showed that recommendation algorithms generate more calibrated recommendations for consistent profiles. As a future work, we aim to further explore the relationship between profile richness and recommendation calibration by taking into account other metrics like profile size and entropy. 

% For bibtex users:
\bibliography{ISMIRtemplate}

\begin{thebibliography}{10}

\bibitem{abdollahpouriWSDM}
Himan Abdollahpouri, Masoud Mansoury, Robin Burke, and Bamshad Mobasher.
\newblock The impact of popularity bias on fairness and calibration in
  recommendation.
\newblock {\em arXiv preprint arXiv:1910.05755}, 2019.

\bibitem{abdollahpourirmse1}
Himan Abdollahpouri, Masoud Mansoury, Robin Burke, and Bamshad Mobasher.
\newblock The unfairness of popularity bias in recommendation.
\newblock {\em In Workshop on Recommendation in Multistakeholder Environments
  (RMSE)}, 2019.

\bibitem{burke2017}
Robin Burke, Nasim Sonboli, Masoud Mansoury, and Aldo Ordoñez-Gauger.
\newblock Balanced neighborhoods for fairness-aware collaborative
  recommendation.
\newblock In {\em RecSys workshop on Fairness, Accountability and Transparency
  in Recommender Systems}, 2017.

\bibitem{ekstrand2018}
Michael~D. Ekstrand, Mucun Tian, Ion~Madrazo Azpiazu, Jennifer~D. Ekstrand,
  Oghenemaro Anuyah, David McNeill, and Maria~Soledad Pera.
\newblock All the cool kids, how do they fit in?: Popularity and demographic
  biases in recommender evaluation and effectiveness.
\newblock In {\em In Conference on Fairness, Accountability and Transparency},
  pages 172--186, 2018.

\bibitem{ghazanfar2014leveraging}
Mustansar~Ali Ghazanfar and Adam Pr{\"u}gel-Bennett.
\newblock Leveraging clustering approaches to solve the gray-sheep users
  problem in recommender systems.
\newblock {\em Expert Systems with Applications}, 41(7):3261--3275, 2014.

\bibitem{Guo2015}
Guibing Guo, Jie Zhang, Zhu Sun, and Neil Yorke-Smith.
\newblock Librec: A java library for recommender systems.
\newblock In {\em UMAP Workshops}, 2015.

\bibitem{lika2014}
Blerina Lika, Kostas Kolomvatsos, and Stathes Hadjiefthymiades.
\newblock Facing the cold start problem in recommender systems.
\newblock {\em Expert Systems with Applications}, 41(4):2065--2073, 2014.

\bibitem{mansoury2018automating}
Masoud Mansoury, Robin Burke, Aldo Ordonez-Gauger, and Xavier Sepulveda.
\newblock Automating recommender systems experimentation with librec-auto.
\newblock In {\em Proceedings of the 12th ACM Conference on Recommender
  Systems}, pages 500--501. ACM, 2018.

\bibitem{mansoury2019biasdisparity}
Masoud Mansoury, Bamshad Mobasher, Robin Burke, and Mykola Pechenizkiy.
\newblock Bias disparity in collaborative recommendation: Algorithmic
  evaluation and comparison.
\newblock In {\em RecSys workshop on Recommendation in Multi-Stakeholder
  Environments}, 2019.

\bibitem{steck2018}
Harald Steck.
\newblock Calibrated recommendations.
\newblock In {\em Proceedings of the 12th ACM Conference on Recommender
  Systems}, pages 154--162. ACM, 2018.

\bibitem{virginia2018}
Virginia Tsintzou, Evaggelia Pitoura, and Panayiotis Tsaparas.
\newblock Bias disparity in recommendation systems.
\newblock {\em CoRR}, abs/1811.01461, 2018.

\bibitem{yao2017}
Sirui Yao and Bert Huang.
\newblock Beyond parity: Fairness objectives for collaborative filtering.
\newblock In {\em In Advances in Neural Information Processing Systems}, pages
  2921--2930, 2017.

\end{thebibliography}

\end{document}